# Stark shift control of single optical centers in diamond

Ph. Tamarat[1], T. Gaebel[1], J. R. Rabeau[2], M. Khan[4], A. D. Greentree[2], H. Wilson[2], L. C. L. Hollenberg[2,3], S. Prawer[2,3], P. Hemmer[4], F. Jelezko[1], J. Wrachtrup[1]

[1] *University of Stuttgart, 3.Physical Institute, 70550 Stuttgart, Germany*

[2] *School of Physics, University of Melbourne, Victoria, Australia*

[3] *Center of Excellence for Quantum Computer Technology, School of Physics, University of Melbourne*

[4] *Department of Electrical and Computer Engineering, Texas A&M University, USA*

**Abstract**

Lifetime limited optical excitation lines of single nitrogen vacancy (NV) defect centers in diamond have been observed at liquid helium temperature. They display unprecedented spectral stability over many seconds and excitation cycles. Spectral tuning of the spin selective optical resonances was performed via the application of an external electric field (i.e. the Stark shift). A rich variety of Stark shifts were observed including linear as well as quadratic components. The ability to tune the excitation lines of single NV centers has potential applications in quantum information processing.

Coupling between light and single spins in solids has attracted widespread attention particularly for applications in quantum computing and quantum communications[1]. The nitrogen-vacancy defect (NV) optical center in diamond is a particularly attractive solid state system for such applications. Its strong optical transition allows photoluminescence-based detection of single defect centers[2]. The potential of the NV center as a single photon source has been well recognized over the past few years[3, 4]. Furthermore, because of its paramagnetic spin ground state, there are applications for quantum memory and quantum repeater systems[5]. In particular the long spin decoherence time (0.35ms), optical control of spin states[6-8] and the robustness of the spin coherence have enabled the demonstration of basic building block for quantum computing even at room temperature[9].

Recently it was demonstrated that the permanent magnetic dipole moment of the NV center can be exploited to couple defects for a separation distance of a few nm. Whilst this demonstrates the capability for the generation of correlated quantum states in defect center clusters, coupling based on this technique will be difficult to scale to many qubit systems. Other coupling schemes have recently been proposed which use instead their optical transition dipole moments and in some cases envisage coupling of the NV center to cavities. At the core of many such schemes is the underlying assumption that the optical transition can be tuned in resonance either with another NV center or with a cavity via an external applied field. Therefore, the ability to tune the frequency of spin-selective optical transitions of single NV centers is of crucial importance for any scalable architecture based on diamond NV centers. Externally controlled magnetic and electric fields are among the most prominent parameters that can be used for such control. Electric fields in particular allow for wide tuning of eigenstates. The electric field induced shift of the optical resonance lines has been observed for single atoms, ions in the gas phase[10] and single molecules[11] and quantum dots[12, 14] in the solid state. By contrast, for color centers in diamond, only a few bulk studies on electric field induced spectral line shifts have been carried out[15]. Usually these studies are difficult because the magnitude of the Stark effect is of the order of the inhomogeneous linewidth. Moreover, ensemble averaging masks the individual Stark behaviors of the defects.

The main challenge until now in using the low temperature optical excitation and emission lines of single defects is mostly related to the spectral diffusion which broadens the linewidths when the observation time exceeds the typical time interval between spectral jumps. Spectral bandwidths exceeding the lifetime-limited (transform-limited) value were reported previously on NV centers[16, 17]. In the present work we demonstrate stable and lifetime-limited optical excitation lines of single NV centers in diamond at low temperature, achieved by careful selection of materials and sample preparation. We took advantage of the high quality factor of the optical resonances to investigate for the first time the Stark effect on single NV centers and to demonstrate that a precise control of the NV center optical transition frequency is possible.

The NV defect in diamond comprises a substitutional nitrogen plus a vacancy in an adjacent lattice site. The defect belongs to the $C_{3v}$ symmetry group. It has an electron spin triplet ground state[18] S=1 with total symmetry $^3A_2$ (see Fig. 1). The transition to the optically excited $^3E$ state (ΔE=1.945eV), usually ascribed to the negatively charged form of the NV center (NV⁻), is a strongly allowed optical dipole transition[19] with a concomitant intense, red-shifted fluorescence. Single NV defect centers in a (100) type IIa diamond were investigated. The defects were imaged with a laboratory built low temperature confocal microscope. The electric field was generated using electrodes deposited on the diamond surface with 50 μm spacing. Electrodes were made by depositing 20nm of Mo and 200nm of Ag onto diamond at about 350 °C. A 400nm overcoat of $SiO_2$ was then deposited on top of the entire surface in order to insulate the electrodes from the helium flow and prevent electrical breakdown.

Fig. 2 shows fluorescence excitation lines of single defects in different samples at T=1.8 K. In nitrogen rich (type Ib) diamonds no narrow excitation lines are found. Upon reducing the nitrogen content the excitation linewidth reduces. The ground state of the substitutional nitrogen impurity is found approximately 1.7 eV below the conduction band of diamond. It is thus optically excited simultaneously with the NV center if there is any nitrogen in the focal volume of the excitation laser. However, because its energy separation to the diamond conduction band is slightly smaller than the $^3A$-$^3E$ transition energy, optical excitation of nitrogen impurities causes photoionisation of the defect. As a consequence fluctuating charges in the conduction band of diamond cause spectral diffusion and dephasing of the NV center excitation line. The inset of Fig. 2 shows the fluorescence excitation line at the zero phonon transition frequency of a single defect center at T=1.8 K in nitrogen pure (type IIa) diamond samples. The line has a lorentzian shape with a linewidth of 13 MHz. Given an excited state lifetime for the NV center of 11.5 ns[20], the measured linewidth is determined by the time-energy uncertainty relation. To the best of our knowledge this is the first time that a transform limited linewidth has been observed for a single NV center in diamond. With a high quality factor of a few $10^7$, and the remarkable spectral stability observed for some defects without any requirements for optical repumping, these excitation lines are well suited for investigation of the Stark effect at the single defect level.

At the second order in the local electric field **F** acting on the defect, the Stark shift of the electronic transition is given by

$$h\Delta\upsilon = -\Delta\boldsymbol{\mu}\cdot\mathbf{F} - \frac{1}{2}\mathbf{F}\cdot\Delta\boldsymbol{\alpha}\cdot\mathbf{F}, \qquad (1)$$

where $\Delta\boldsymbol{\mu}$ and $\Delta\boldsymbol{\alpha}$ are respectively the change in dipole moment and the change of polarizability tensor between the excited and ground states. Higher order terms are expected to arise in eq. 1 via higher order hyperpolarizabilities[21]. However, estimation of the cubic term, related to the change in the second order polarizability $\chi^{(2)}$, indicates that it should be smaller than the quadratic term by a factor $F/F_{int}$, where $F_{int}$ is the internal field acting within the defect center. Since the applied electric fields range up to a few MVm$^{-1}$ in our measurements, third and higher order corrections are neglected in the Stark analysis. In order to extract the dipole moment and polarizability changes from the data, it is necessary to determine the value of the local field F from the externally applied E. Here we adopt the simplest approach using the Lorentz local field approximation $F = E(\varepsilon + 2)/3$, where ε is the dielectric constant for diamond.

Fig. 3 shows a series of excitation spectra recorded while changing step wise the applied electric field from zero to 0.32 MVm$^{-1}$. The resulting spectral trails correspond to individual defects which are all sampled within the focused laser spot. These trails give an overview of the diversity of the individual Stark behaviors which have been previously unobservable in ensemble measurements. For electric fields up to a few MVm$^{-1}$, the Stark shift of the optical resonance is well fitted by linear and quadratic dependences. Comparing this expression to eq. (1) leads to the values of the change in the dipole moment $\Delta\mu$ and the change in polarizability $\Delta\alpha$ along the local field **F** direction. Fig. 4 exemplifies three different cases of Stark behavior. Fig. 4a describes a defect center showing a linear Stark effect with a slope of -6.3 GHz/(MVm$^{-1}$) which corresponds to $\Delta\mu$ = 1.3 D (1 D = 3.33 10$^{-30}$ Cm). This value is very similar to values found in other cases, e.g. aromatic molecules in solid molecular hosts[11]. Most defect centers also show a quadratic electric field dependence. Fig. 4b represents a defect with an almost exclusively quadratic component with $\Delta\alpha$ = -3.5 10$^4$ Å$^3$ and Fig. 4c depicts a case where both a linear and a quadratic component are present. Note that we have used here the polarizability volume which is the polarizability divided by $4\pi\varepsilon_o$. A statistical investigation of a few tens of single NV centers gave a distribution for $\Delta\mu$ between -1.5 D and +1.5 D. The values of $\Delta\alpha$ were found between -6 10$^4$ Å$^3$ and 0. It should be noted that for quantum dots[12](+ ref 15) and molecules[11] mostly positive values for $\Delta\alpha$ are found, since excited states are generally more polarizable than ground states. Apparently, this is not the case for the NV center.

Our experimental findings are remarkable for two reasons. Firstly, for a non-centrosymmetric defect one would expect to find a predominantly linear Stark shift. As apparent from Fig. 3 this is not generally the case in the present experiments. In fact, some defects do not show any linear shift whilst most of them reveal a strong quadratic Stark component which is due to the polarizability of the system. The polarizability is related to the dielectric constant $\varepsilon$ via $\alpha = (\varepsilon - 1)v\varepsilon_0$ where v is the defect center volume. In this simple classical model, changes in the polarizability $\Delta\alpha$ are expected to be of the order of v. In small systems like single defects or molecules values of $\Delta\alpha$ of around ~10 Å$^3$ are usually found[11]. Stark shift experiments on shallow acceptor levels of impurity levels in diamond on the other hand yield polarizability values which are one order of magnitude larger than the ones found in the present case[22]. Since for the case of the acceptors the Stark shift is quadratic in the Bohr radius of the exciton[23], we conclude that the delocalization of the electron-hole wavefunction of the NV center is approximately three times smaller than that for shallow acceptor levels, e.g. for boron in diamond where it is around 3nm. From electron spin resonance studies[24] and quantum chemical calculations[25] it is clear that the NV wavefunction is mostly localized on the dangling bond of the three carbon atoms. However, 30% of the electron density is delocalized over the shell of next nearest neighbours which might yield an effective Bohr radius of a size necessary to explain the polarizabilities found. Whether a defect shows a predominantly linear or quadratic component depends on the degeneracy of the excited state, which is an orbital doublet. Those systems with degenerate levels will reveal a predominant linear Stark shift. Those defects where this degeneracy is lifted by e.g. strain or possibly the orientation of the center with respect to the field show a substantial quadratic contribution. Preliminary DFT calculations carried out on the defect support this argument. Calculations of the electron density for the HOMO and LUMO states in a uniform electric field indicate an orientation dependent breaking of the electronic symmetry of the defect which may lead to the appearance of a quadratic Stark effect for particular field orientations. Precise numerical values of the linear shift and polarizability, however, require an accurate estimate of the excited state structure which in case of the NV center is difficult to ascertain reliably in these calculations.

The second striking feature is the decrease in fluorescence intensity which was observed for some excitation lines out of a given applied field range, as seen in the inset of Fig. 1. This observation was not found to be related to an electric field induced charge transfer process of the NV center. This was checked by monitoring the fluorescence spectrum of single defects excited at 532 nm. No signal arising from the neutral configuration of the NV center, which has a zero phonon line at 575 nm, appeared under applied field. Therefore the disappearance of the excitation lines is tentatively attributed to field enhanced non-radiative or spin cross relaxation decay channels. This may be related to a lifting of the degeneracy of the excited $^3E$ state and a concomitant state mixing which is explained in more detail as follows. It is known from defects in the $C_{3v}$ symmetry group that an externally applied field with components orthogonal to the three fold symmetry axis of the defects, which is defined to be along z, lifts the degeneracy of the excited state $^3E$ level[15]. In the absence of an external field, the $^3E$ state dominant splitting is either caused by spin-spin or axial $\lambda L_z S_z$ spin-orbit interaction[26]. Neither of the two interactions causes mixing of the $S_z$ and $S_{x,y}$ sublevels and thus causes a change in spin projection upon optical excitation. With an electric field parallel to the three fold symmetry axis, both E orbital doublets undergo an identical shift proportional to the field strength F. This will only influence the line spectral position. In the situation where components of F are orthogonal to z, the $^3E$ sublevels split as $\pm(F_x^2+F_y^2)^{1/2}$ and get mixed. This might cause spin cross relaxation or enhanced inter-system crossing with a concomitant change in fluorescence intensity. The relevant energy scale for this to happen is the strength of the spin orbit coupling, i.e. 30 GHz[27]. The electric fields used in our experiment were able to shift the resonances by a comparable value. Hence an electric field induced state mixing is an effective mechanism for fluorescence reduction.

In conclusion, we have demonstrated that stable and Fourier-transform limited optical excitation lines were found in diamond at low temperature when the sample material was carefully chosen. This allowed the first investigation of the Stark effect at the single defect level, revealing complex behaviors which are hidden in ensemble measurements. Due to the degeneracy of the excited $^3E$ level and the small spin orbit coupling, the NV center shows a rich variety of Stark shifts with widely distributed linear and quadratic components. It should be pointed out that spectral tuning of the spin selective optical transition by an external electric field is a first step toward controlled optical dipole coupling between close defect centers. The application of finer nanometer sized gates and possibly feedback control of the applied voltage are future developments which pave the way toward optical control of the NV center ground state spins, for example in defect center pairs[28] with potential applications in quantum memory and quantum repeater schemes.

**Acknowledgement**

This work has been supported by the European Commission under the Integrated Project Qubit Applications (QAP) funded by the IST directorate as Contract Number 015848 and by the DFG via the SFB/TR 21 as well as the Landesstiftung BW via the program "Atomoptik". The study was initiated as part of a DARPA QuIST effort, AFOSR contract C02-00060. Ph. Tamarat acknowledges the Alexander von Humboldt foundation for a research fellowship.

**Figure captions**

Fig. 1

Energy level scheme of the NV center in diamond. The optically excited $^3E$ state is doubly degenerated. The inset shows the fluorescence intensity of a single defect as a function of an applied electric field. Points are measured values and the solid curve is a guide to the eye.

Fig. 2

Fluorescence excitation zero phonon line of single NV defect centers at T = 1.8 K. (a), (b) Excitation line for a single defect in a diamond nanocrystal (type Ib). Spectra (a) and (b) have been recorded with and without deshelving 488 nm laser, respectively. (c) Excitation line in type IIa crystal. The excitation power was 5 nW, far below the saturation value of about 100 nW. The line shown in the inset is fitted with a lorentzian profile and has a lifetime-limited homogeneous width of 13 MHz.

Fig. 3

Spectral trails of different single defects found in the same laser spot. Excitation spectra have been recorded with step wise changing of the applied field from 0 to 0.32 MVm$^{-1}$. Some excitation lines are observed only for a specific range of field values.

Fig. 4

Examples for the Stark shift of three different defect centers. The local electric field was calculated from the externally applied bias using the Lorentz local field correction (for details see text). (a) Linear Stark effect with $\Delta\mu$ = 1.3 D. (b) Mostly quadratic Stark effect with $\Delta\mu$ = -37 mD and $\Delta\alpha$ = -3.5 10$^4$ Å$^3$. (c) Linear and quadratic Stark dependences, with $\Delta\mu$ = 1.1 D and $\Delta\alpha$ = -5.4 10$^4$ Å$^3$.

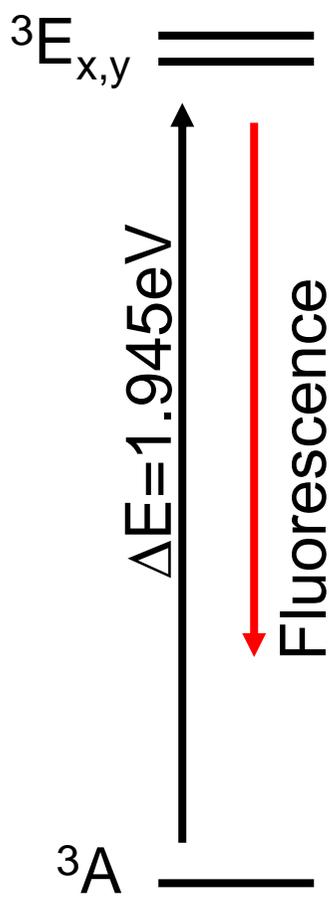
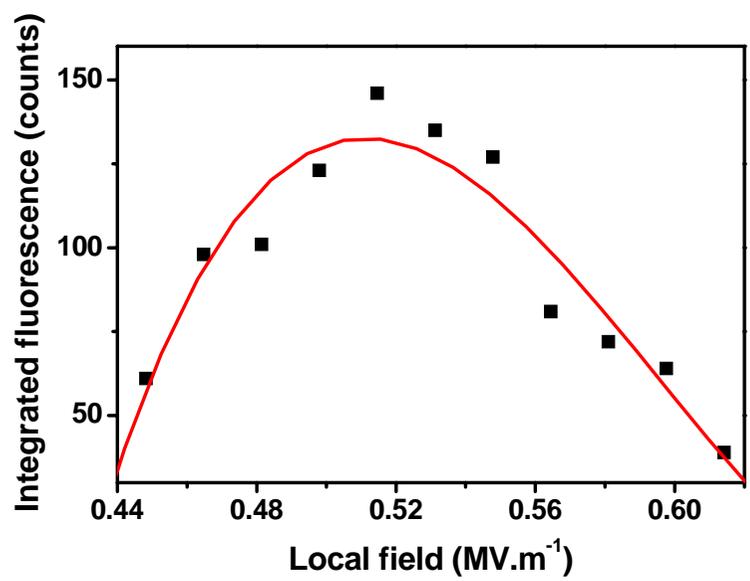

Tamarat et al. Fig. 1

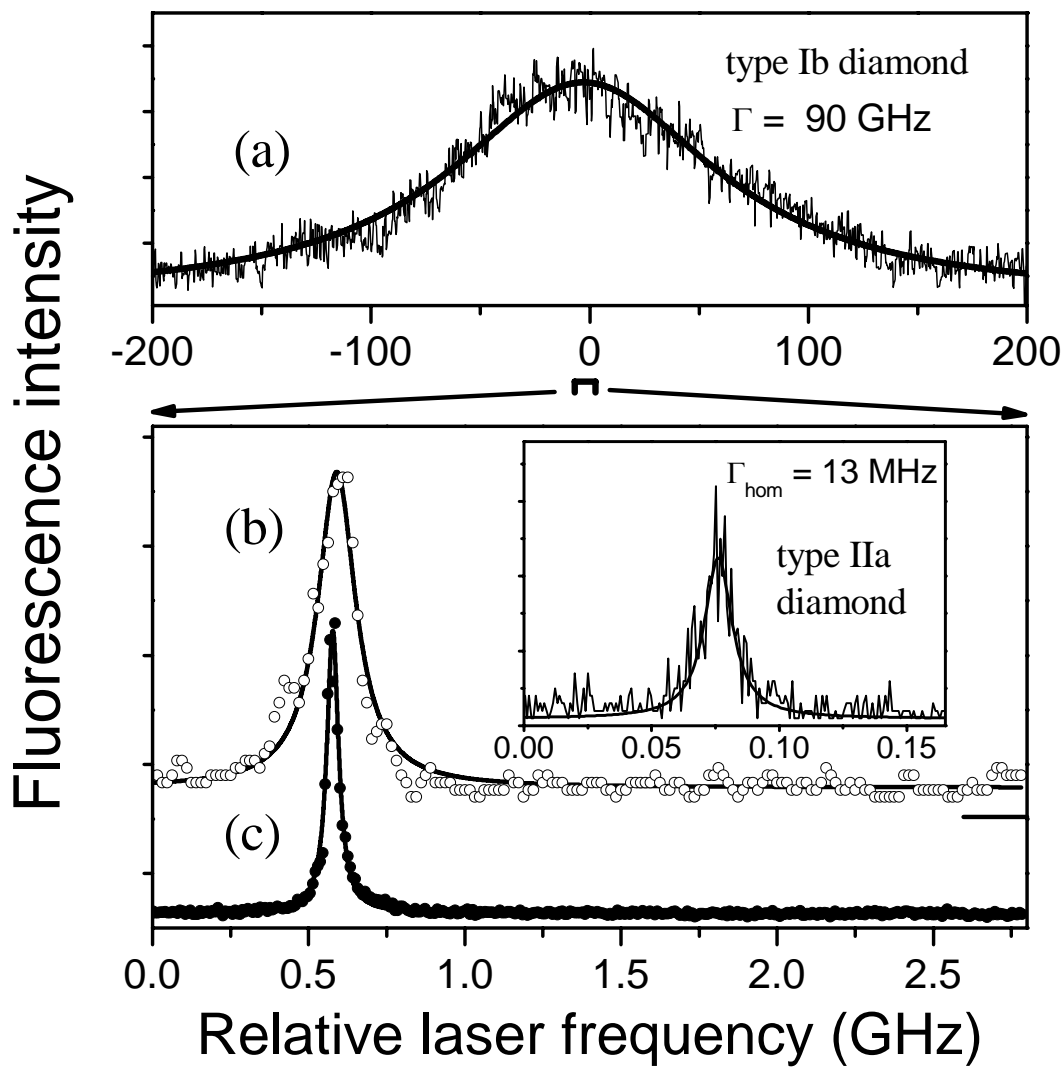

Tamarat et al. Fig. 2

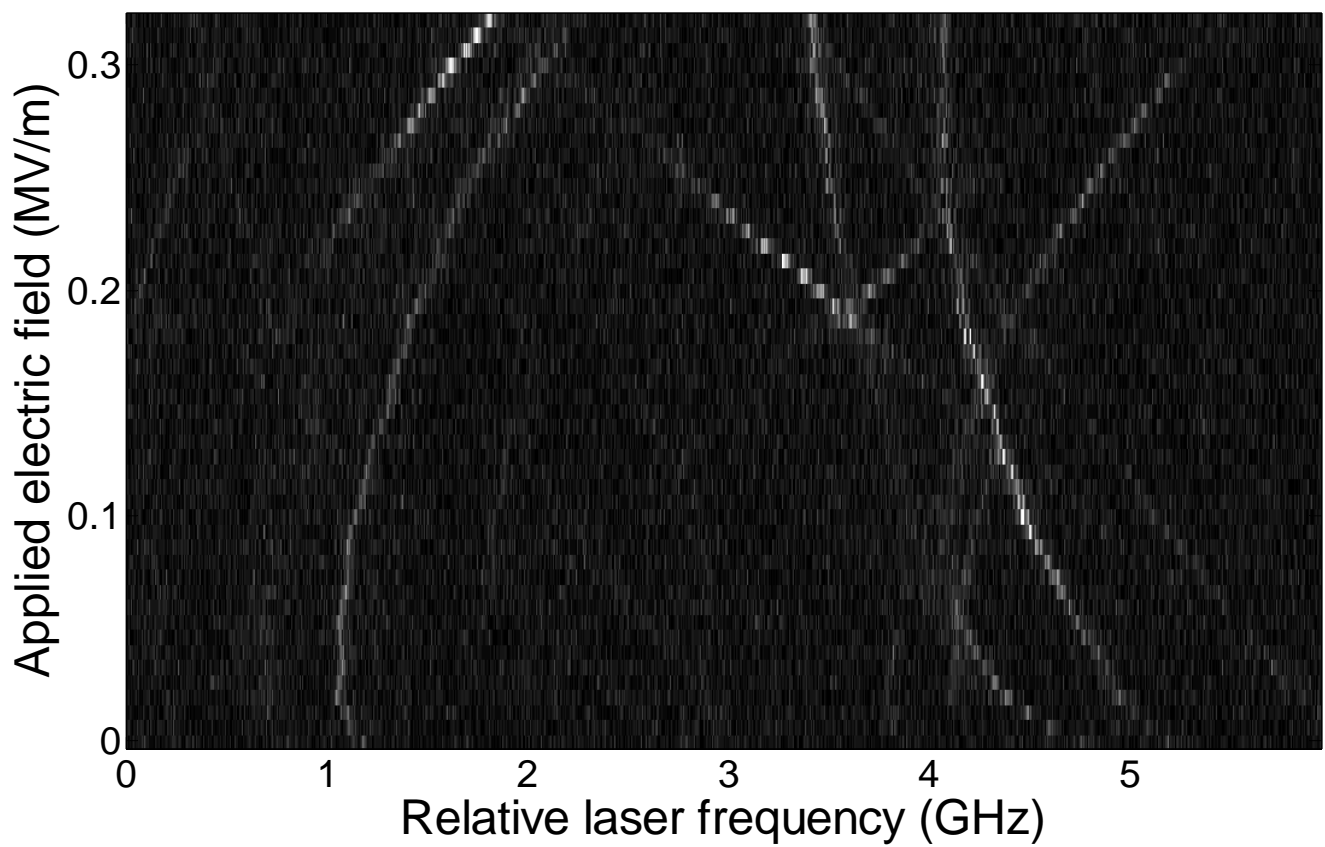

Tamarat et al. Fig. 3

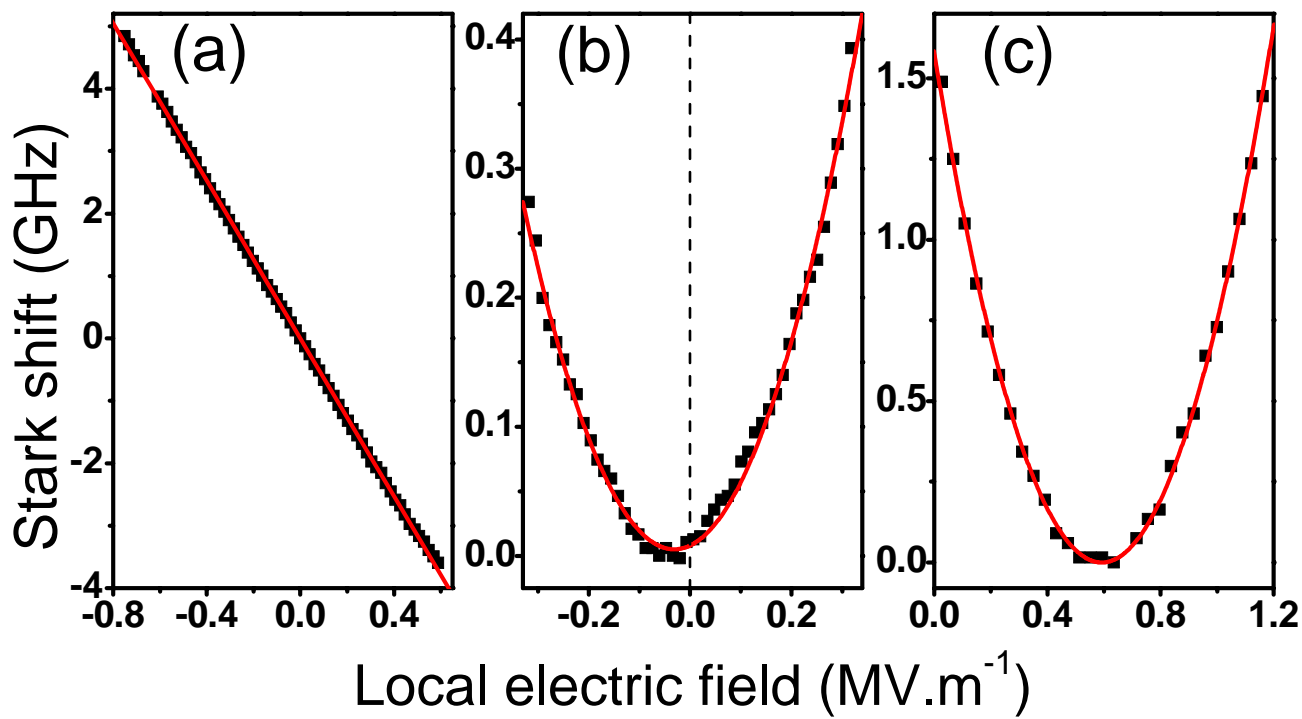

Tamarat et al. Fig. 4